\documentclass{article}
\usepackage{spconf,amsmath,graphicx}
\usepackage{amsfonts}
\usepackage{multirow}

\title{EMOTIONAL VOICE CONVERSION WITH CYCLE-CONSISTENT \\ ADVERSARIAL NETWORK}
\name{\begin{tabular}{c}Songxiang Liu, Yuewen Cao, Helen Meng\end{tabular}}
\address{
   Human-Computer Communications Laboratory \\
   Department of Systems Engineering and Engineering Management,\\
  The Chinese University of Hong Kong, Shatin, N.T., Hong Kong SAR, China \\
  \small \tt \{sxliu, ywcao, hmmeng\}@se.cuhk.edu.hk
  }

\begin{document}
\ninept
\maketitle
\begin{abstract}
Emotional Voice Conversion, or emotional VC, is a technique of converting speech from one emotion state into another one, keeping the basic linguistic information and speaker identity. Previous approaches for emotional VC need parallel data and use dynamic time warping (DTW) method to temporally align the source-target speech parameters. These approaches often define a minimum generation loss as the objective function, such as L1 or L2 loss, to learn model parameters. Recently, cycle-consistent generative adversarial networks (CycleGAN) have been used successfully for non-parallel VC.  This paper investigates the efficacy of using CycleGAN for emotional VC tasks. Rather than attempting to learn a mapping between parallel training data using a frame-to-frame minimum generation loss, the CycleGAN uses two discriminators and one classifier to guide the learning process, where the discriminators aim to differentiate between the natural and converted speech and the classifier aims to classify the underlying emotion from the natural and converted speech. The training process of the CycleGAN models randomly pairs source-target speech parameters, without any temporal alignment operation. The objective and subjective evaluation results confirm the effectiveness of using CycleGAN models for emotional VC. The non-parallel training for a CycleGAN indicates its potential for non-parallel emotional VC.

\end{abstract}

\section{Introduction}
\label{sec:intro}

Human speech is a complex signal that contains rich information, which includes linguistic information, para- and non-linguistic information.
While linguistic information and para-linguistic information controlled by the speaker help make the expression convey information precisely, non-linguistic information such as emotion accompanied with speech plays an important role in human social interaction. Compared with Voice Conversion (VC) \cite{abe1990voice, shikano1991speaker, liu2018voice, liu2019jointly}, which is a technique to convert one speaker's voice to sound like that of another, emotional Voice Conversion, or emotional VC, is a technique of converting speech from one emotion state into another one, keeping the basic linguistic information and speaker identity.

Most of the VC approaches focus more on the conversion of short-time spectral features, but less on the conversion of the prosody features such as F0 \cite{toda2007voice, sun2015voice, nakashika2016non, hsu2017voice, liu2018voice}. 
In these works, F0 features are usually converted by a simple logarithmic Gaussian (LG) normalized transform. For emotional VC, however, parametrically modeling the prosody features is important, since the prosody plays an important role in conveying various types of non-linguistic information, such as intention, attitude and mood, which represent the emotions of the speaker \cite{luo2017emotional}. 
Recently, there has been tremendous active research in modeling prosodic features of speech for emotional VC, most of which involve modeling two prosodic elements, namely the F0 contour and energy contour. 
A Gaussian mixture model (GMM) and a classification regression tree model were adopted to model the F0 contour conversion from neutral speech to emotional speech in \cite{tao2006prosody}. 
A system for transforming the emotion in speech was built in \cite{inanoglu2007system}, where the F0 contour was modeled and generated by context-sensitive syllable-based HMMs, the duration was transformed using phone-based relative decision trees, and the spectrum was converted using a GMM-based or a codebook selection approach. 

Prosody is inherently supra-segmental and hierarchical in nature, of which the conversion is affected by both short- and long-term dependencies, such as the sequence of segments, syllables, words within an utterance as well as lexical and syntactic systems of a language \cite{xu2011speech, hirst1998intonation, yu2012review, wennerstrom2001music, kawasaki1997alternatives}. There have been many attempts to model prosodic characteristics in multiple temporal levels, such as the phone, syllable and phrase levels \cite{latorre2008multilevel, wu2010hierarchical, obin2011stylization, qian2011improved}. 
Continuous wavelet transform (CWT) can effectively model F0 in different temporal scales and significantly improve speech synthesis performance \cite{suni2013wavelets}. The CWT methods were also adopted for emotional VC. CWT was adopted for F0 modeling within the non-negative matrix factorization (NMF) model \cite{ming2015fundamental}, and for F0 and energy contour modeling within a deep bidirectional LSTM (DBLSTM) model \cite{ming2016deep}. Using CWT method to decompose the F0 in different scales has also been explored in \cite{luo2017emotional, luo2016emotional}, where neural networks (NNs) or deep belief networks (DBNs) were adopted.

While previous work has shown the efficacy of using GMMs, DBLSTMs, NNs and DBNs to model the feature mapping for the spectral and the prosodic features, they all need parallel data and parallel training, which means the source and target data should have parallel scripts and a dynamic time warping (DTW) method is used to temporally align the source and target features before training the models. Parallel training data is more difficult to collect than non-parallel data in many cases. Besides, the use of DTW may introduce alignment errors, which degrades VC performance.
Moreover, previous emotional VC approaches often define a minimum generation loss as the objective function, such as L1 or L2 loss. One of the issues using a minimum generation loss is an over-smoothing effect often observed in the generated speech parameters. Since this loss may also be inconsistent with human's perception of speech, directly optimizing the model parameters using a minimum generation loss may not generate speech that sounds natural to human. Generative adversarial networks (GANs) \cite{goodfellow2014generative} have been incorporated into TTS and VC systems \cite{saito2018statistical}, where it is found that GAN models are capable of generating more natural spectral parameters and F0 than conventional minimum generation error training algorithm regardless of its hyper-parameter settings. 
Since any utterance spoken by a speaker with the source or target emotional state can be used as training sample, if a non-parallel emotional VC model can achieve comparable performance with the parallel counterparts, it will be more flexible, more practical and more valuable than parallel emotional VC systems. The recently emerged cycle-consistent generative adversarial network (CycleGAN) \cite{zhu2017unpaired}, which belongs to the large family of GAN models, provides a potential way to achieve non-parallel emotional VC. The CycleGAN was originally designed to transform styles in images, where the styles of the images are translated while the textures remain unchanged. CycleGAN models have been used successfully for developing non-parallel VC systems \cite{kaneko2017parallel, fang2018high}.

In this paper, we investigate the efficacy of using CycleGAN models for emotional VC tasks. Emotional VC is similar to image style transformation, where we can regard the underlying linguistic information as analogous to image content and the accompanying emotion as analogous to image style. Rather than attempting to learn a mapping between parallel training data using a frame-to-frame minimum generation loss, in this paper, the CycleGAN uses two discriminators and one classifier to guide the learning process--the discriminators aim to differentiate between natural and converted speech and the classifier aims to classify the underlying emotion from the natural and converted speech. The spectral features, F0 contour and energy contour are simultaneously converted by the CycleGAN model. We utilize CWT or logarithmic representation of the F0 and energy features. Although the training data we use is parallel, a non-parallel training process is adopted to learn the CycleGAN model, which means that source and target features are randomly paired during training, without any temporal alignment process. 
The objective and subjective evaluation results confirm the effectiveness of using CycleGAN models for emotional VC. The advantages offered by the CycleGAN model include (i)utilizing GAN loss instead of minimum generation loss, (ii)getting rid of source-target alignment errors and (iii) flexible non-parallel training, etc.
The non-parallel training for a CycleGAN indicates its potential for non-parallel emotional VC.

The rest of this paper is organized as follows: Section 2 introduces the CycleGAN model for emotional VC and Section 3 describes the details of implementation. Section 4 gives the experimental setups and evaluations. Conclusions are drawn in Section 5.

\section{EMOTIONAL VC WITH CYCLEGAN}
\label{sec:pagestyle}

\begin{figure}[th]
  \centering
  \centerline{\includegraphics[width=7.8cm]{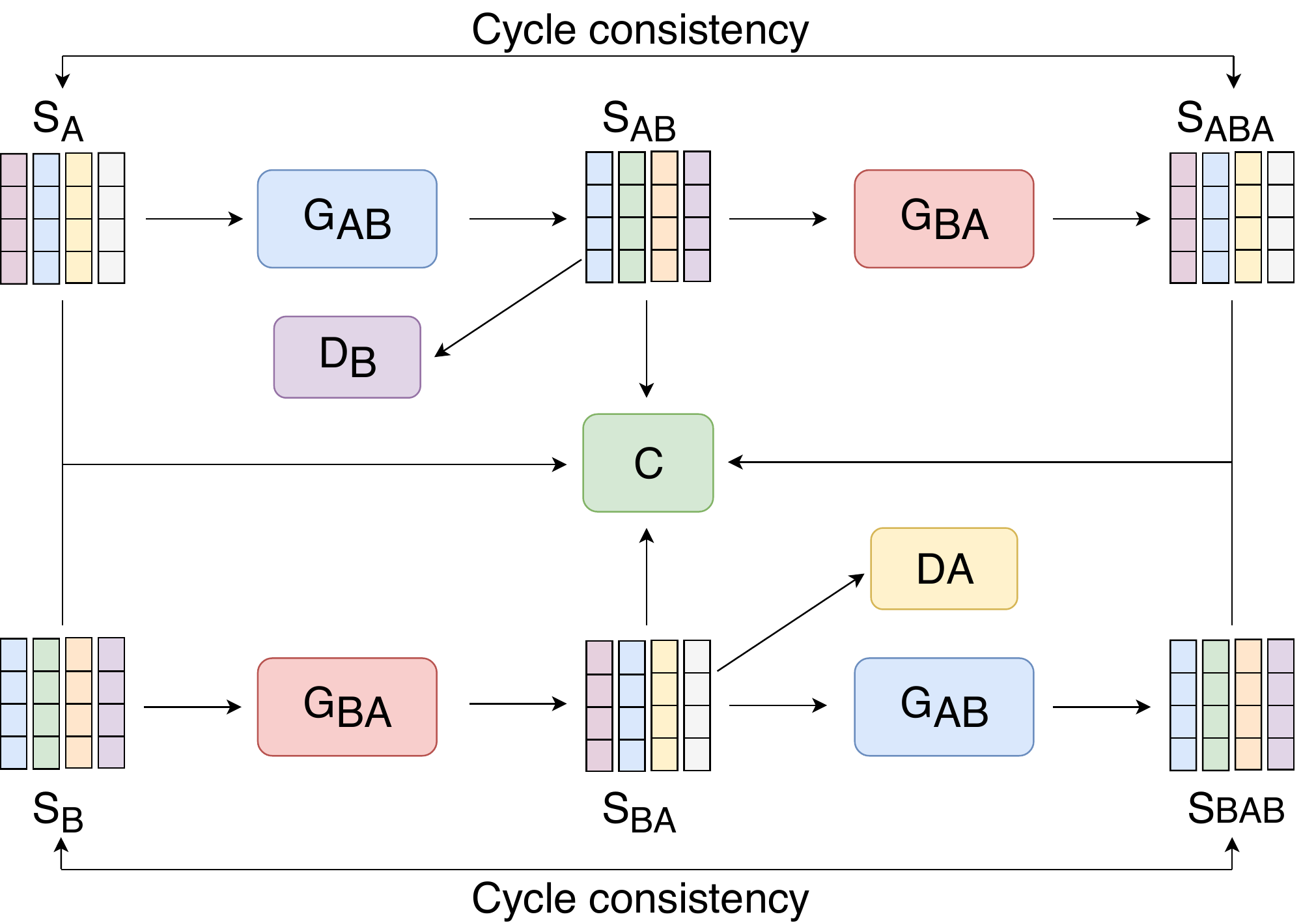}}
  \caption{CycleGAN structure for emotional voice conversion. $G_{AB}$ and $G_{BA}$ refer to generators. $D_A$ and $D_B$ refer to discriminators. $C$ refers to emotion classifier. $S$ denotes genuine or converted spectral and prosodic features.}
  \label{fig:cycleganfig}
\end{figure}

The CycleGAN model consists of two generators ($G_{AB}$ and $G_{BA}$), two discriminators ($D_A$ and $D_B$) and one emotion classifier (C), as shown in Fig.~\ref{fig:cycleganfig}, where we denote spectral and prosodic features in the domain of emotion $A$ as $S_A$, spectral and prosodic features in the domain of emotion $B$ as $S_B$, respectively. 
$S_{AB}$ is the converted spectral and prosodic features from emotion $A$ to emotion $B$ by the generator $G_{AB}$, while $S_{ABA}$ is the features converted back to emotion $A$ by the generator $G_{BA}$ from $S_{AB}$. 
To effectively learn parameters of the generators, discriminators and classifier, several losses are defined as follows.



{\bf{Adversarial Loss:}} Generator $G_{AB}$ serves as a mapping function from emotion domain $A$ to emotion domain $B$, while generator $G_{BA}$ do the opposite, serving as a mapping function from emotion domain $B$ to emotion domain $A$. The discriminators, $D_A$ and $D_B$, aim to distinguish between genuine and converted spectral and prosodic features, i.e., discriminator $D_A$ distinguishes between $S_A$ and $S_{BA}$, and discriminator $D_B$ distinguishes between $S_B$ and $S_{AB}$.
To this end, an adversarial loss, which measures how distinguishable the converted features $S_{AB}$ from the genuine target domain features $S_B$, is defined as

\begin{equation}
  \begin{aligned}
  \mathcal{L}_{adv}(&G_{AB}, D_B) = \mathbb{E}_{S_B\sim p_{data}(S_B)}[\log D_B(S_B)] \\
                                 &+ \mathbb{E}_{S_A\sim p_{data}(S_A)}[\log (1 - D_B(G_{AB}(S_A))]. \\
  \end{aligned}
\end{equation}

The adversarial loss distinguishing the converted features $S_{BA}$ from the genuine source domain features $S_A$ has a similar formulation. 
%


{\bf{Cycle Consistency Loss:}} The adversarial loss makes $S_A$ and $S_{BA}$ or $S_B$ and $S_{AB}$ as similar as possible while the cycle consistency loss guarantees that an input $S_A$ can retain its original form after passing through the two generators $G_{AB}$ and $G_{BA}$. Using the notation in Figure \ref{fig:cycleganfig}, $S_{ABA}$, which equals to $G_{BA}(G_{AB}(S_A))$, should not diverge too much from $S_A$. 
This is very important for emotional voice conversion, since we do not want to change the linguistic and speaker information during the conversion process. 
The cycle consistency loss is defined as 

\begin{equation}
  \begin{aligned}
  \mathcal{L}_{cyc}(&G_{AB}, G_{BA}) \\
   &=\mathbb{E}_{S_A \sim p_{data}(S_A)}[||G_{BA}(G_{AB}(S_A)) - S_A||_1]\\
                                    &+\mathbb{E}_{S_B \sim p_{data}(S_B)}[||G_{AB}(G_{BA}(S_B)) - S_B||_1],
  \end{aligned}
\end{equation}

where $||\cdot||_1$ means L1 norm.


{\bf{Emotion Classification Loss:}} To explicitly guide the CycleGAN model to learn the emotion conversion function, we add additional emotion classification loss to the original model. Specifically, an accompanying emotion classifier $C$ as shown in Fig.~\ref{fig:cycleganfig} is trained, which determines whether $S_A$, $S_{BA}$ and $S_{ABA}$ match the desired emotion label $A$, as well as whether $S_B$, $S_{AB}$ and $S_{BAB}$ match the desired emotion label $B$. To achieve this, the following emotion classification loss is introduced:

\begin{equation} \label{eq:3}
  \begin{aligned}
  \mathcal{L}_{emo}(G_{AB}, G_{BA}, C) &= \mathcal{L}_{emo-A}(G_{AB}, G_{BA}, C) \\
    &+\mathcal{L}_{emo-B}(G_{AB}, G_{BA}, C),
  \end{aligned}
\end{equation}
where
\begin{equation} \label{eq:4}
  \begin{aligned}
  \mathcal{L}_{emo-A}&(G_{AB}, G_{BA}, C) = d(C(S_A), label_A) \\ 
                          & + d(C(S_{AB}), label_B) + d(C(S_{ABA}), label_A),
  \end{aligned}
\end{equation}
and 
\begin{equation} \label{eq:5}
  \begin{aligned}
  \mathcal{L}_{emo-B}&(G_{AB}, G_{BA}, C) = d(C(S_B), label_B) \\ 
                          & + d(C(S_{BA}), label_A) + d(C(S_{BAB}), label_B).
  \end{aligned}
\end{equation}

In equation (\ref{eq:4}) and (\ref{eq:5}), $d(\cdot)$ can be any divergence function used for classification problem, e.g., the binary cross-entropy loss function. 


{\bf{Full Objective:}} Combining the above adversarial loss, cycle consistency loss and emotion classification loss, the full training objective is:
\begin{equation}
  \begin{aligned}
  \mathcal{L}(G_{AB}, G_{BA}, D_A, D_B, C)  &= \mathcal{L}_{adv}(G_{AB}, D_B)\\
                                            &+ \mathcal{L}_{adv}(G_{BA}, D_A)\\
                                            &+ \lambda_1\mathcal{L}_{cyc}(G_{AB}, G_{BA})\\
                                            &+ \lambda_2\mathcal{L}_{emo}(G_{AB}, G_{BA}, C),
  \end{aligned}
\end{equation}
where $\lambda_1$ and $\lambda_2$ are trade-off parameters, adjusting the relative weights of these loss terms.

\section{IMPLEMENTATION}

In this paper, speech features including Mel-cepstral coefficients (MCCs) and F0 are computed using WORLD \cite{morise2016world}. The spectral features for conversion are Mel-cepstral coefficients (MCCs), which have dimension of 36. The energy contour and the F0 contour, as well as their corresponding CWT decomposition, are computed as in \cite{ming2016deep}. We use the CWT or logarithmic representation for the F0 and energy features. For convenience, we denote the CWT representation of F0 and energy contour as $\mathbf{LF0}_{cwt}$ and $\mathbf{LE}_{cwt}$, respectively, while denote the logarithmic F0 as $\mathbf{LF0}$. Network architecture of the generators, discriminators and classifier are shown in Table \ref{table:1}. The DBLSTM models, the baseline, are set to have the same network architecture as in \cite{ming2016deep}. The hyper-parameters and training details are made available here\footnote{Source code: https://github.com/liusongxiang/CycleGAN-EmoVC} but left out for space limitation.

\begin{table}[ht]
  \caption{\label{table:1} {Network architecture of generators, discriminators and classifier. IN is for instance normalization \cite{ulyanovinstance}. LReLU indicates leakyReLU. ConvTran refers to transpose convolution. $C_{in}$ is the input channel to the layer. $k_1$ and $k_2$ are the input feature height and width divided by 16.
  }}
  \centerline{
  \footnotesize
\begin{tabular}{|c|l|}
\hline
\multicolumn{2}{|c|}{\bf{Generator}}  \\
\hline
          Conv block                        & Conv@3$\times$9$\times$1$\times$64, IN, ReLU                         \\
\hline
\multirow{2}{*}{Down-sampling block}        & Conv@4$\times$8$\times$64$\times$128, stride=2, IN, ReLU             \\
                                            & Conv@4$\times$8$\times$128$\times$256, stride=2, IN, ReLU            \\
\hline
\multirow{2}{*}{Residual block $\times$6}   & Conv@3$\times$3$\times$256$\times$256, IN, ReLU                      \\
                                            & Conv@3$\times$3$\times$256$\times$256, IN                            \\
\hline
\multirow{2}{*}{Up-sampling block}          & ConvTran@4$\times$4$\times$256$\times$128, stride=2, IN, ReLU        \\
                                            & ConvTran@4$\times$4$\times$128$\times$64, stride=2, IN, ReLU         \\
\hline
        Output layer                        & Conv@7$\times$7$\times$64$\times$1                                   \\
\hline 
\hline
\multicolumn{2}{|c|}{\bf{Discriminator/Classifier}}  \\
\hline
        Conv block                          & Conv@4$\times$4$\times$1$\times$64, stride=2, LReLU                  \\
\hline
        Stride block $\times$4              & Conv@4$\times$4${\times}C_{in}{\times}2C_{in}$, stride=2, LReLU      \\
\hline
        Output layer                        & Conv@$k_1{\times}k_2{\times}1024{\times}$1                           \\
\hline 
\end{tabular}}
\end{table}

\begin{table}[th]
\caption{\label{table:2} {Mapping models and features. }}
\centerline{
  \small
\begin{tabular}{|c|c|c|}
\hline
Models  & Converted Features  \\
\hline 
DBLSTM-1 & $\mathbf{MCCs}$   \\
DBLSTM-2 & $\mathbf{MCCs}$, $\mathbf{LF0}$   \\
DBLSTM-3 & $\mathbf{MCCs}$, $\mathbf{LF0}_{cwt}$   \\
DBLSTM-4 & $\mathbf{MCCs}$, $\mathbf{LF0}_{cwt}$, $\mathbf{LE}_{cwt}$   \\
CycleGAN-1 & $\mathbf{MCCs}$   \\
CycleGAN-2 & $\mathbf{MCCs}$, $\mathbf{LF0}$   \\
CycleGAN-3 & $\mathbf{MCCs}$, $\mathbf{LF0}_{cwt}$   \\
CycleGAN-4 & $\mathbf{MCCs}$, $\mathbf{LF0}_{cwt}$, $\mathbf{LE}_{cwt}$   \\
\hline
\end{tabular}}
\end{table}

In the training and conversion stages, MCC features and prosodic features are concatenated, so the model maps these features together. The features are normalized to zero mean and unit variance before being fed into the CycleGAN and DBLSTM models. During conversion, the aperiodicity component remains intact and is directly copied over. 
We first compute the logarithm-scale F0 and energy contour from the converted CWT-represented F0 and energy features, respectively. Then a mean-variance denormalization and an exponential operation are adopted to compute the linear-scale F0 and energy contours of the target emotion from the normalized logarithm-scale ones. If we denote the converted spectral feature as $\mathbf{SP}^c$, which is computed from the converted $\mathbf{MCC}$, and the linear-scale energy contour as $\mathbf{e}^c$, the final converted spectral $\tilde{\mathbf{SP}}^c$ is computed as follow: (i) Compute the energy contour $\mathbf{e}^t$ of $\mathbf{SP}^c$. (ii) Compute the element-wise ratio $\mathbf{r} = \mathbf{e}^c / \mathbf{e}^t$. (iii) Scale the $i$-th frame vector $\mathbf{SP}_i^c$ by $\mathbf{r}_i$ to obtain $\tilde{\mathbf{SP}}_i^c$.


\section{EXPERIMENTS AND RESULTS}
\label{sec:typestyle}

\subsection{Experiment conditions}
We use the CASIA Chinese Emotional Corpus, recorded by Institute of Automation, Chinese Academy of Sciences, where each sentence with the same semantic texts is spoken by 2 female and 2 male speakers in six different emotional tones: happy, sad, angry, surprise, fear, and neutral. 
We choose three emotions (sad, neutral and angry), which form a strong contrast, from one female speaker. We use 260 utterances for each emotion as training set, 20 utterances as validation set and another 20 utterances as evaluation set. 

Note that although the training data is parallel, the training process of the CycleGAN models randomly pairs source-target features, thus dynamic time-warping (DTW) alignment is not needed. Since the DBLSTM models in nature need frame-to-frame mapping between the source-target features, a DTW process is necessary to temporally align the source-target spectral features as well as the prosodic features, i.e., F0 and energy representations. 
The experimental setups are listed in Table \ref{table:2}, where each model does two conversion tasks, which are neutral-to-sad conversion and neutral-to-angry conversion.

\begin{table}[ht]
  \caption{\label{table:3} {MCD and LogF0-MSE results}}
  \center
  \small
\begin{tabular}{|c|c|c|c|c|}
\hline  
\multirow{2}{*}{} & \multicolumn{2}{c|}{MCD (dB)} & \multicolumn{2}{c|}{LogF0-MSE} \\
                  \cline{2-5}
                  & Sad           & Angry        & Sad           & Angry         \\
                  \hline
Source            & 10.87         & 14.56        & 0.063         & 0.098         \\
DBLSTM-1          & 9.97          & 10.59        & 0.065         & 0.132         \\
DBLSTM-2          & \bf{9.55}     & \bf{9.74}    & 0.027         & \bf{0.039}    \\
DBLSTM-3          & 10.60         & 11.38        & 0.029         & 0.045         \\
DBLSTM-4          & 10.57         & 11.83        & \bf{0.025}    & 0.042         \\
CycleGAN-1        & 10.43         & 10.60        & 0.065         & 0.132         \\
CycleGAN-2        & 10.70         & 10.49        & 0.030         & 0.057         \\
CycleGAN-3        & 10.04         & 10.55        & 0.030         & 0.075         \\
CycleGAN-4        & 10.30         & 10.26        & 0.034         & 0.059         \\
\hline
\end{tabular}
\end{table}

\subsection{Objective evaluation}

The Mel Cepstral Distortion (MCD) is used for the objective evaluation of spectral conversion. The MCD is computed as:


\begin{equation}
  MCD  =  \frac{10}{\ln 10} \sqrt{2\sum_{i=1}^{36}(\mathbf{MCC}_i^t - \mathbf{MCC}_i^c)^2},
\end{equation}
where $\mathbf{MCC}_i^t$ and $\mathbf{MCC}_i^c$ represent the target and the converted Mel-cepstral, respectively.
The LogF0 mean squared error (MSE) is computed to evaluate the F0 conversion, which has the form

\begin{equation}
  MSE = \frac{1}{N} \sum_{i=1}^{N}(\log(F0_i^t) - \log(F0_i^c))^2,
\end{equation}
where $F0_i^t$ and $F0_i^c$ denote the target and the converted F0 features, respectively. The average MCD and LogF0-MSE results are illustrated in Table \ref{table:3}. The MCD and LogF0-MSE between the source and the target emotion are computed as reference.

Based on the MCD results, the best performing approach is DBLSTM-2, which converts spectral features and logarithmic F0 contour. CycleGAN-2 has the worst MCD for neutral-to-sad conversion, while DBLSTM-4 has the worst MCD for neutral-to-angry. Comparing CycleGAN-3 with DBLSTM-3 and CycleGAN-4 with DBLSTM-4, we see that the CycleGANs have lower MCDs than their DBLSTMs counterparts for both conversions, although there is no explicitly defined minimum generation loss when training CycleGANs. 
Based on the LogF0-MSE results, DBLSTM-4 has the lowest for the sad emotion and DBLSTM-2 has the lowest for the angry emotion.
Comparing DBLSTM-1 to DBLSTM-(2-4) and CycleGAN-1 to CycleGAN-(2-4), we can see that simultaneously modeling F0 features (CWT or logarithmic representations) with spectral features achieves better conversion result than just using a simple logarithmic Gaussian normalized transform for F0 conversion.   
It is reasonable that the DBLSTMs achieve lowest results for both MCD and LF0-MSE metrics, since they are trained by optimizing the explicitly defined minimum generation loss between the DTW-aligned source and target speech features, and the MCD computation also uses DTW to align the converted and the genuine speech features. 


\subsection{Subjective Evaluation}

\begin{table}[ht]
  \caption{\label{table:4} {Subjective classification results}}
  \small
   \center
\begin{tabular}{|c|c|c|c|c|}
\hline 
\multicolumn{2}{|c|}{Target \textbackslash Perception} & Sad & Angry & Neutral \\
\hline
\multirow{2}{*}{DBLSTM-1}   & Sad      & {\bf{45.2}}\%     & 12.3\%        & 42.5\%         \\
                            \cline{2-5}
                            & Angry    & 2.5\%      & {\bf{83.8}}\%       & 13.7\%         \\
                            \hline
\multirow{2}{*}{CycleGAN-1} & Sad      & 43.8\%     & 0.8\%        & {\bf{55.4}}\%         \\
                            \cline{2-5}
                            & Angry    & 3.3\%      & {\bf{78.9}}\%       & 17.8\%         \\
                            \hline
\multirow{2}{*}{CycleGAN-2} & Sad      & {\bf{65.6}}\%     & 2.2\%        & 32.2\%         \\
                            \cline{2-5}
                            & Angry    & 6.7\%      & {\bf{62.5}}\%       & 30.8\%         \\
                            \hline
\multirow{2}{*}{CycleGAN-3} & Sad      & {\bf{56.7}}\%     & 2.1\%        & 41.2\%         \\
                            \cline{2-5}
                            & Angry    & 3.3\%     & {\bf{62.3}}\%       & 34.4\%         \\
                            \hline
\multirow{2}{*}{CycleGAN-4} & Sad      & {\bf{56.9}}\%     & 0.9\%        & 42.2\%         \\
                            \cline{2-5}
                            & Angry    & 1.2\%      & {\bf{74.4}}\%       & 24.4\%         \\
\hline 
\end{tabular}
\end{table}

A subjective emotion classification test is conducted, where each model has 20 testing utterances (10 for each conversion). The listeners are ask to label the stimuli as more 'sad' or more 'angry' when compared with a neutral reference. 16 listeners take part in this test. Since the converted waveforms by DBLSTM-(2-4) are obviously worse in speech naturalness than those by other settings according to the preliminary listening test, we only conduct subjective evaluations for five models, i.e., DBLSTM-1, CycleGAN-(1-4). The subjective classification results are shown in Table \ref{table:4}. We see some inconsistency between the objective metrics and the subjective evaluation results, which is often encountered in VC and TTS literatures.

For the conversion from neutral to sad, CycleGAN-2, which converts spectral features together with the logarithmic F0 simultaneously, achieves the best result. For the conversion from neutral to angry, DBLSTM-1 achieves the best result, while CycleGAN-1 also achieves good result with degradation by only 5.8\%. The neutral-to-sad conversion has lower results than the neutral-to-angry conversion under both DBLSTM and CycleGAN settings except CycleGAN-2. One possible reason is that perception of emotional state from sad speech is more difficult than that of angry speech when using the neutral as reference. 
Sad speech is characterized by low energy and slow speech rate, while angry speech is characterized by high energy and fast speech rate. Comparing the different CycleGAN settings, which convert different speech features, we can see that different feature combinations obtain good results for different emotion conversion, where CycleGAN-1 has high result for neutral-to-angry conversion and CycleGAN-2 has high result for neutral-to-sad conversion. The subjective emotion classification test shows the effectiveness to use CycleGAN model for emotional VC. Since the training process of CycleGANs is non-parallel, where source-target speech parameters are randomly paired, this work validates the utility of the CycleGAN approach in emotional VC based on training with non-parallel emotional databases.

\section{CONCLUSIONs}
\label{sec:majhead}

This paper investigates the efficacy of using CycleGAN for emotional VC tasks. Rather than attempting to learn a mapping between parallel training data using a frame-to-frame minimum generation loss, the CycleGAN uses two discriminators and one classifier to guide the learning process, where the discriminators aim to differentiate between the natural and converted speech and the classifier aims to classify the underlying emotion from the natural and converted speech. The training process of the CycleGAN models randomly pairs source-target speech parameters, thus DTW process is not needed. The objective and subjective evaluation results confirm the effectiveness of using CycleGAN models for emotional VC. To sum up, the advantages offered by the CycleGAN model include (i)utilizing GAN loss instead of minimum generation loss, (ii)getting rid of source-target alignment errors and (iii) flexible non-parallel training, etc. The non-parallel training process also indicates the potential to use non-parallel emotional speech data for developing emotional VC systems, which will be our future work.




\vfill\pagebreak



\bibliographystyle{IEEEbib}
\small
\bibliography{strings,refs}

\end{document}